\documentclass[%
 reprint,
 amsmath,amssymb,
 aps,
pre
]{revtex4-1}

\usepackage{graphicx}
\usepackage{dcolumn}
\usepackage{bm}
\usepackage{amsmath}
\usepackage{amssymb}
\newcommand{\beq}{\begin{equation}}
\newcommand{\eeq}{\end{equation}}

\begin{document}

\preprint{APS/123-QED}

\title{Localization of eigenvector centrality in networks with a cut vertex}

\author{Kieran J. Sharkey}
\email{kjs@liverpool.ac.uk}
 \affiliation{Department of Mathematical Sciences, University of Liverpool, Liverpool L69 7ZL, United Kingdom}


\begin{abstract}
We show that eigenvector centrality exhibits localization phenomena on networks that can be easily partitioned by the removal of a vertex cut set, the most extreme example being networks with a cut vertex. Three distinct types of localization are identified in these structures. One is related to the well-established hub node localization phenomenon and the other two are introduced and characterized here. We gain insights into these problems by deriving the relationship between eigenvector centrality and Katz centrality. This leads to an interpretation of the principal eigenvector as an approximation to more robust centrality measures which exist in the full span of an eigenbasis of the adjacency matrix.
\end{abstract}

\pacs{Valid PACS appear here}
\maketitle

\section{Introduction}
\label{1}
Cataloging individual nodes and the connections between them forms the underlying data in many areas of science and technology, such as the world-wide-web, social networks, biochemical pathways, transportation networks, and power grids \cite{Dorogovtsev_mendes}\cite{Newman03}\cite{Newman}. This underlying concept of a graph or network is the same across disciplines and it is not surprising that the same issues emerge. A common problem is how to identify which nodes are most significant. This is valuable if we wish to identify the most important pages on the internet or the most influential people from social network analysis or to target resources at controlling an epidemic on a network of contacts. 

Measures of node importance are often termed `centrality' \cite{Newman03}\cite{Newman}\cite{Freeman}. Degree centrality is the most obvious measure of the relative importance of nodes and refers to how many nearest neighbors that any given node has. In general, the structure of a network can be represented by an adjacency matrix $A$ such that element $A_{ij}=1$ if node $j$ is connected towards node $i$ and $A_{ij}=0$, otherwise. For an $n$ by $n$ adjacency matrix $A$ representing an undirected network, degree centrality is given by
\beq
\textbf{d}=A\textbf{1},
\label{degree}
\eeq
where $\textbf{1}$ is the appropriate column vector of ones.

One of the main deficiencies of degree centrality is that a simple tally of the number of neighbors does not account for whether those neighbors are themselves important. Generally it is reasonable to suppose that nodes with high centrality should confer a higher centrality onto their neighbors than lower centrality nodes.
A standard method for resolving this problem is eigenvector centrality \cite{Bonacich}, which relates to the eigenvalue equation for $A$:
\beq
\textbf{u}=\frac{1}{\mu}A\textbf{u}.
\label{ev_eq}
\eeq
Comparing with Eq.~(\ref{degree}), the eigenvalue equation has the required form; instead of summing over the number of neighboring nodes 
with equal weight, we have a weighted sum where each neighbor contributes centrality in proportion to its own centrality $\textbf{u}$. For this equation to have a solution, it is of course required that $\mu$ is an eigenvalue of $A$ and that $\textbf{u}$ is its corresponding eigenvector. 

To avoid unnecessary complication, let us suppose that $A$ is strongly connected (irreducible) and let us also assume that it is undirected. From the Perron-Frobenius theorem, the principal eigenvalue of $A$ has a corresponding eigenvector in its eigenspace whose components are all positive. Consequently, this is the solution of Eq.~(\ref{ev_eq}) that is generally used to define eigenvector centrality. 

To set the scene for what follows, we note that while positive values are likely to be a desirable attribute for a centrality measure and we should also expect that the principal eigenvector contains more information than any of the other eigenvectors, this is not a sufficient reason to neglect the other eigenvectors. Essential ranking information could also exist in the direction of some of the other eigenvectors and so we are not able to guarantee that eigenvector centrality will always give a sensible ranking of node importance.

Indeed, problems with eigenvector centrality known as localization have been observed whereby the centrality is localized on just a few nodes in the network. This is particularly apparent when networks have highly connected hub nodes \cite{Goh}\cite{Farkas}\cite{Goltsev}\cite{Martin}\cite{Pastor}, but also occurs when networks have high modularity \cite{Eriksen}. Here we develop a more detailed understanding of localization phenomena in scenarios where networks are easily partitionable. We use the term localization here to refer to any unreasonable focusing of centrality on parts of the network.

We initially consider networks with a cut vertex (Sec.~\ref{2}) since the main results can be fully explored in this simpler scenario (Sec.~\ref{3}). We then show how this extends to the general case of networks partitioned by an arbitrary vertex cut set (Sec.~\ref{4}) and define three types of localization (Sec.~\ref{5}). In Sec.~\ref{6} we provide an interpretation of eigenvector centrality as an approximation to more robust centrality measures such as Katz centrality.

\section{Eigenvector centrality for networks with a cut vertex}
\label{2}
We first define a network which is partitionable by the removal of a single vertex. Consider a network with adjacency matrix $A$ and a cut vertex such that its removal results in $m$ disconnected components (or partitions) with adjacency matrices $P_i$ of order $p_i$ by $p_i$ for $i\in\{1,2,\dots,m\}$. The adjacency matrix $A$ has the form
\beq
A=
\left (
\begin{array}{ccccc}
P_1 & 0_{p_1\times p_2} & \cdots & 0_{p_1\times p_m} & \textbf{b}_1     \\
0_{p_2\times p_1} & P_2  & \cdots & 0_{p_2\times p_m} &\textbf{b}_2     \\
\vdots & \vdots &  \ddots & \vdots & \vdots \\
0_{p_m\times p_1} & 0_{p_m\times p_2}  & \cdots & P_m & \textbf{b}_m \\
\textbf{b}^T_1 & \textbf{b}^T_2  &\cdots &\textbf{b}^T_m & 0
\end{array}
\right ).
\label{matrix}
\eeq
Here, the notation $0_{p_i\times p_j}$ denotes the $p_i$ by $p_j$ zero matrix and the column vector $\textbf{b}_i$ of length $p_i$ describes connections from the cut vertex to partition $P_i$. Since $A$ is strongly connected and undirected, it follows that each partition $P_i$ is also strongly connected.

The form of eigenvector centrality for this network can be obtained from the eigenvalue equation (see also Martin {\it et al}. \cite{Martin} which effectively considered $m=1$). Suppose that the principal eigenvalue of $A$ is $\mu$ and that the corresponding eigenvector is 
\beq
\textbf{u}=\left (
\begin{array}{c}
\textbf{x}_1 \\
\textbf{x}_2 \\
\vdots \\
\textbf{x}_m \\
v
\end{array}
\right )
\label{EV_struct}
\eeq
where $\textbf{x}_i$ are column vectors of length $p_i$ and $v$ is a scalar. Substituting this and Eq.~(\ref{matrix}) into Eq.~(\ref{ev_eq}) gives
\beq
P_i\textbf{x}_i+v\textbf{b}_i=\mu\textbf{x}_i
\nonumber
\eeq
for $i\in\{1,2,\dots,m\}$. Solving this for $\textbf{x}_i$ gives
\beq
\textbf{x}_i=\frac{v}{\mu}\left (I-\frac{1}{\mu}P_i\right )^{-1}\textbf{b}_i,
\nonumber
\eeq
where $I$ is the appropriately sized identity matrix. By substituting these values into Eq.~(\ref{EV_struct}) we obtain
\beq
\textbf{u}\propto\left (
\begin{array}{c}
M_1\textbf{b}_1 \\
M_2\textbf{b}_2 \\
\vdots \\
M_m\textbf{b}_m \\
\mu
\end{array}
\right ),
\label{EV_cent}
\eeq
where
\beq
M_i=\left (I-\frac{1}{\mu}P_i\right )^{-1}.
\label{M_defA}
\eeq

To investigate this further, it is valuable to develop a leading eigenvector approximation to Eq.~(\ref{EV_cent}). Since $A$ is undirected, we can assume an orthonormal eigenbasis for each of the $P_i$ given by the vectors $\textbf{w}_i^1,\textbf{w}_i^2,\dots,\textbf{w}_i^{p_i}$ with corresponding eigenvalues $\lambda_i^1,\lambda_i^2,\dots,\lambda_i^{p_i}$. We can write the vector $\textbf{b}_i$ in the corresponding basis:
\beq
\textbf{b}_i=g_i^1\textbf{w}_i^1+g_i^2\textbf{w}_i^2+\dots+g_i^{p_i}\textbf{w}_i^{p_i}
\nonumber
\eeq
with coordinates given by the projection of $\textbf{b}_i$ onto the relevant basis vectors: $g_i^j=\textbf{b}_i\cdot\textbf{w}_i^j$ for $j\in\{1,2,\dots, p_i\}$. Additionally, decomposing the inverse matrix Eq.~(\ref{M_defA}) as a power series in $P_i/\mu$,
\beq
M_i=I+\frac{P_i}{\mu}+\frac{P_i^2}{\mu^2} +\dots,
\nonumber
\eeq
now lets us write 
\begin{eqnarray}\nonumber
M_i\textbf{b}_i &=&(I+\frac{P_i}{\mu}+\frac{P_i^2}{\mu^2}+\dots)(g_i^1\textbf{w}_i^1+g_i^2\textbf{w}_i^2+\dots+g_i^{p_i}\textbf{w}_i^{p_i}) \\ \nonumber
&=& g_i^1\textbf{w}_i^1+g_i^2\textbf{w}_i^2+\dots+g_i^{p_i}\textbf{w}_i^{p_i} \\ \nonumber
&&+g_i^1\frac{\lambda_i^1}{\mu}\textbf{w}_i^1+g_i^2\frac{\lambda_i^2}{\mu}\textbf{w}_i^2+\dots+g_i^{p_i}\frac{\lambda_i^{p_i}}{\mu}\textbf{w}_i^{p_i} \\ \nonumber
&&+g_i^1\frac{(\lambda_i^1)^2}{\mu^2}\textbf{w}_i^1+g_i^2\frac{(\lambda_i^2)^2}{\mu^2}\textbf{w}_i^2+\dots+g_i^{p_i}\frac{(\lambda_i^{p_i})^2}{\mu^2}\textbf{w}_i^{p_i} \\ \nonumber
&& + \dots \\
&=&\frac{g^1_i\textbf{w}^1_i}{1-\lambda_i^1/\mu}+\frac{g^2_i\textbf{w}^2_i}{1-\lambda^2_i/\mu}+ \dots + \frac{g^{p_i}_i\textbf{w}^{p_i}_i}{1-\lambda^{p_i}_i/\mu}.
\label{M_expansion}
\end{eqnarray} 
It is worth clarifying that since $A$ is irreducible and $P_i$ is a subgraph of $A$, $\mu$ is larger in modulus than the eigenvalues of $P_i$ (\cite{Grantmacher}, pp. 83-84). 

To keep the notation simple, let us denote the leading eigenvector of partition $P_i$ by $\textbf{w}_i$ and its associated eigenvalue by $\lambda_i$. We can now make a leading eigenvector approximation $\tilde{\textbf{u}}$ to capture the main characteristics of Eq.~(\ref{EV_cent}) in most circumstances:
\beq
\tilde{\textbf{u}}\propto\left (
\begin{array}{c}
\textbf{w}_1(\textbf{b}_1\cdot\textbf{w}_1)/(\mu-\lambda_1) \\
\textbf{w}_2(\textbf{b}_2\cdot\textbf{w}_2)/(\mu-\lambda_2) \\
\vdots \\
\textbf{w}_m(\textbf{b}_m\cdot\textbf{w}_m)/(\mu-\lambda_m) \\
1
\end{array}
\right ).
\label{EV_approx}
\eeq
\section{Localization in networks with a cut vertex}
\label{3}
We can gain insights into the limitations of eigenvector centrality from Eq.~(\ref{EV_cent}) as well as from its approximation Eq.~(\ref{EV_approx}). We will do this by making a detailed analysis of some numerical evaluations of eigenvector centrality that exhibit localization. First we can place the known results on hub node localization into the current context.

Consider a class of network described by Eq.~(\ref{matrix}) with a cut vertex that results in $m$ partitions. Suppose that its spectral radius $\mu(n)$ scales with size such that $\lim_{n\rightarrow\infty}\lambda_i(n)/\mu(n)= 0$ where $\lambda_i(n)$ is the spectral radius of partition $P_i$. It then follows from Eqs.~(\ref{EV_cent}) and~(\ref{M_defA}) that the the eigenvector centrality of the nodes in $P_i$ tends toward $\textbf{b}_i$ and becomes uninformative. In some cases this can be problematic for centrality; the $m=1$ case corresponds to hub node localization where an unreasonable focusing of centrality on the hub node and its immediate neighbors can occur. This has been observed on several networks \cite{Goh}\cite{Farkas}\cite{Goltsev}\cite{Pastor} and established as a phase transition on a class of undirected random graphs \cite{Martin}. 

The form of Eq.~(\ref{EV_cent}) results in other types of localization and these are the main focus of this work. In particular, notice that the eigenvector centrality of nodes in subgraph $P_i$ is directly dependent on the nodes, defined by $\textbf{b}_i$, that the cut vertex connects to. This suggests that there could be a large nonlocal impact on the entire subgraph of the choice of connecting nodes. 

To explore this in detail it is valuable to make a comparison of the average centrality in two specific partitions; we shall consider $P_1$ and $P_2$. From the approximation of eigenvector centrality Eq.~(\ref{EV_approx}), the ratio of the average centrality in subgraph $P_1$ to the average centrality in subgraph $P_2$ is approximated by
\beq
\rho=\frac{p_2(\textbf{1}\cdot\textbf{w}_1)}{p_1(\textbf{1}\cdot\textbf{w}_2)}\frac{(\textbf{b}_1\cdot\textbf{w}_1)}{(\textbf{b}_2\cdot \textbf{w}_2)}\frac{(\mu-\lambda_2)}{(\mu-\lambda_1)}.
\label{ratio}
\eeq
The first factor can be shown to be bounded between $\sqrt{p_2}/p_1$ and $p_2/\sqrt{p_1}$ by making use of the bounds $\|\textbf{x}\|_2\le\|\textbf{x}\|_1\le\sqrt{p}\|\textbf{x}\|_2$ provided by the $l_2$-norm on the $l_1$-norm for a vector $\textbf{x}$ of dimension $p$. For subgraphs of similar size and type, its value is typically close to 1. 

\begin{figure}
\centerline{\includegraphics[width=.5\textwidth]{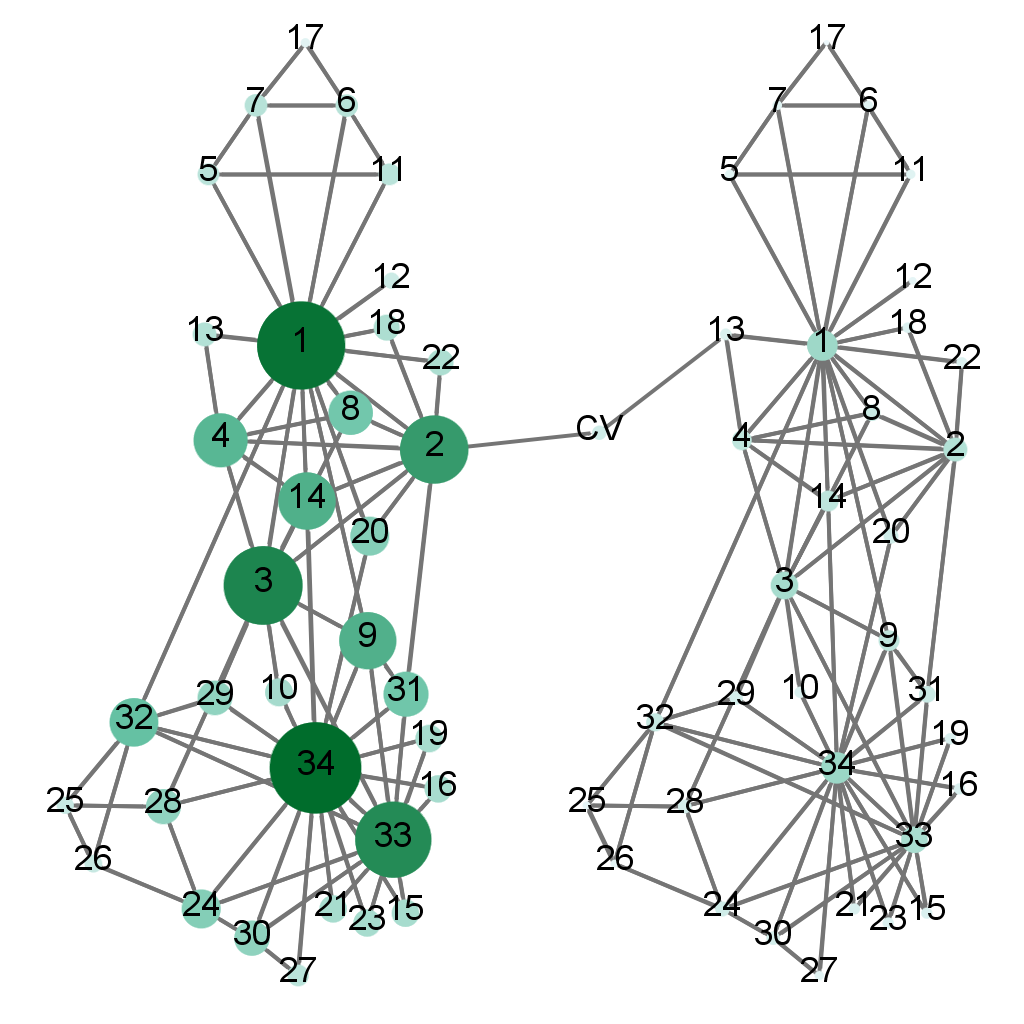}}
\caption{The classic karate club network \cite{Zachary} is duplicated. Node 2 in the left network is connected to node 13 in the network on the right via an additional node (CV) which is a cut vertex. The size of the nodes increases with their eigenvector centrality and the color from white through to green is also changing with increasing eigenvector centrality.}
\label{Karate_twice}
\end{figure}

We now identify two types of localization, one associated with the second factor and one associated with the third factor in Eq.~(\ref{ratio}). To investigate the second factor, it is informative to consider the situation where the two partitions $P_1$ and $P_2$ are isomorphic so that $\textbf{w}_1=\textbf{w}_2$, $p_1=p_2$ and $\lambda_1=\lambda_2$ and so we are left with just the second factor: $\rho=(\textbf{b}_1\cdot\textbf{w}_1)/(\textbf{b}_2\cdot\textbf{w}_1)$. In addition to describing the ratio between the average centralities of the partitions, in this particular case it also gives the ratio between corresponding nodes. An example of such a network is shown in Fig.~\ref{Karate_twice} where the classic karate club network of Zachary \cite{Zachary} 
has been duplicated and then linked by an additional connecting node of degree two (the cut vertex). The only difference between the duplicated subgraphs is that the cut vertex connects to node 2 in the left subgraph and to node 13 in the right subgraph.

Identifying the left subgraph with $P_1$ and the right subgraph with $P_2$, we obtain $\rho=3.157$, rounded to four significant figures. Here, $\rho$ reduces to the ratio of the eigenvector centrality of node 2 to the eigenvector centrality of node 13 when computed on the original (single) karate club network. When computing the thirty-four individual ratios of the eigenvector centrality of node $i$ in the left subgraph to the eigenvector centrality of node $i$ in the right subgraph for $i\in\{1,2,\dots,34\}$, we find that, when excluding corresponding pairs of nodes 2 and 13, this has mean 3.157 rounded to four significant figures with standard deviation 0.015. Corresponding nodes 2 differ in eigenvector centrality by the ratio 3.236 and corresponding nodes 13 differ by the ratio 2.502 reflecting differences due to the connection of these nodes to the cut vertex which would be described by including the other terms in Eq.~(\ref{M_expansion}). The small node-specific variation described by the standard deviation also relates to the contributions from the other terms in Eq.~(\ref{M_expansion}) which are neglected in Eq.~(\ref{ratio}). These terms are small because the values $\mu=6.738$ and $\lambda_1=\lambda_2=6.726$ cause the terms with the largest eigenvalue to be far more significant.

For most applications we would expect a useful centrality measure to provide more or less the same centrality values to the corresponding nodes in the two subgraphs except for some deviation near to the connecting nodes. However, a network-wide impact of the choice of connecting nodes is observed whereby the centralities in the left subgraph are significantly more than those in the right subgraph. By changing the choice of connecting nodes, a nonlocal network-wide impact on every node occurs as the value of the ratio $\rho$ changes.

\begin{figure}
\centerline{\includegraphics[width=.5\textwidth]{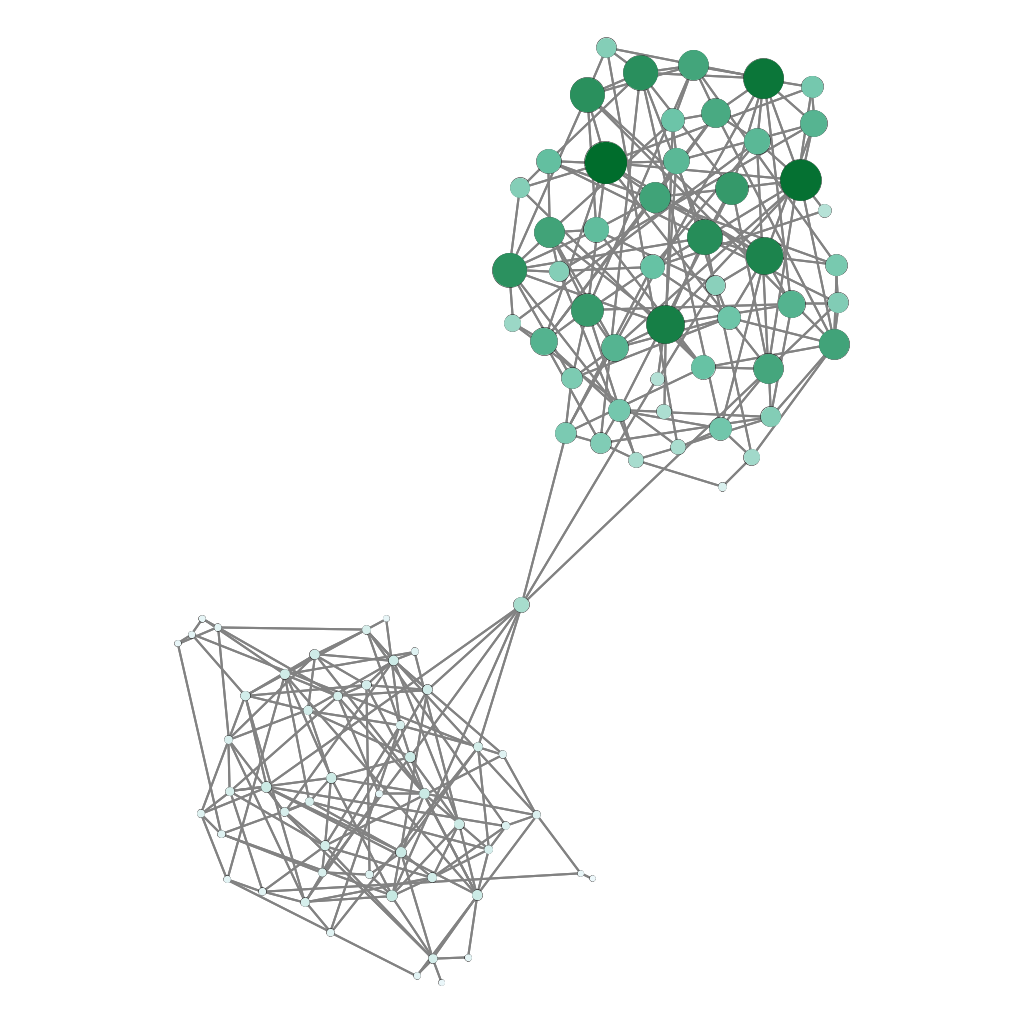}}
\caption{A network formed by connecting two different fifty-node Erd\H os-R\' enyi random graphs together with a cut vertex. The bottom graph has average degree 5.84 and the top has average degree 6.32. The cut vertex connects to five nodes in the bottom subgraph and to three nodes in the top subgraph. The size of nodes increases with their eigenvector centrality and their color also changes from white through to green.}
\label{random}
\end{figure}

When subgraphs $P_1$ and $P_2$ are different, there are contributions from all factors in Eq.~(\ref{ratio}).  The last factor depends on the principal eigenvalues of subgraphs $P_1$ and $P_2$. We should  expect some dependence, but this term can be very sensitive to whether $\lambda_1$ or $\lambda_2$ is closest to $\mu$, and consequently $\rho$ can be very large or small because of this. Additionally, as in the previous example, we also have dependence on the second factor describing the nonlocal impact of the choice of connecting nodes. Figure~\ref{random} illustrates two different Erd\H os-R\' enyi random graphs of the same order and similar density joined together. Here most of the centrality is in the top-right subgraph, demonstrating localization behavior. The average eigenvector centrality of nodes in the upper subgraph is found to be 7.301 times greater than those in the lower subgraph. This is captured by Eq.~(\ref{ratio}) which gives $\rho=7.312$. The first factor in Eq.~(\ref{ratio}) has value 1.024. The second has value 0.3917 which partly reflects the fact that the lower graph has five connections and the upper one has three, leading to a higher amount of the eigenvector centrality of the lower subgraph being directly connected to the cut vertex than the eigenvector centrality of the upper one. The third factor is 18.23 illustrating sensitivity to the principal eigenvalues which have values $\mu=7.190$, $\lambda_1=7.171$ and $\lambda_2=6.856$ and highlighting another cause of localization.

\begin{figure}
\centerline{\includegraphics[width=.55\textwidth]{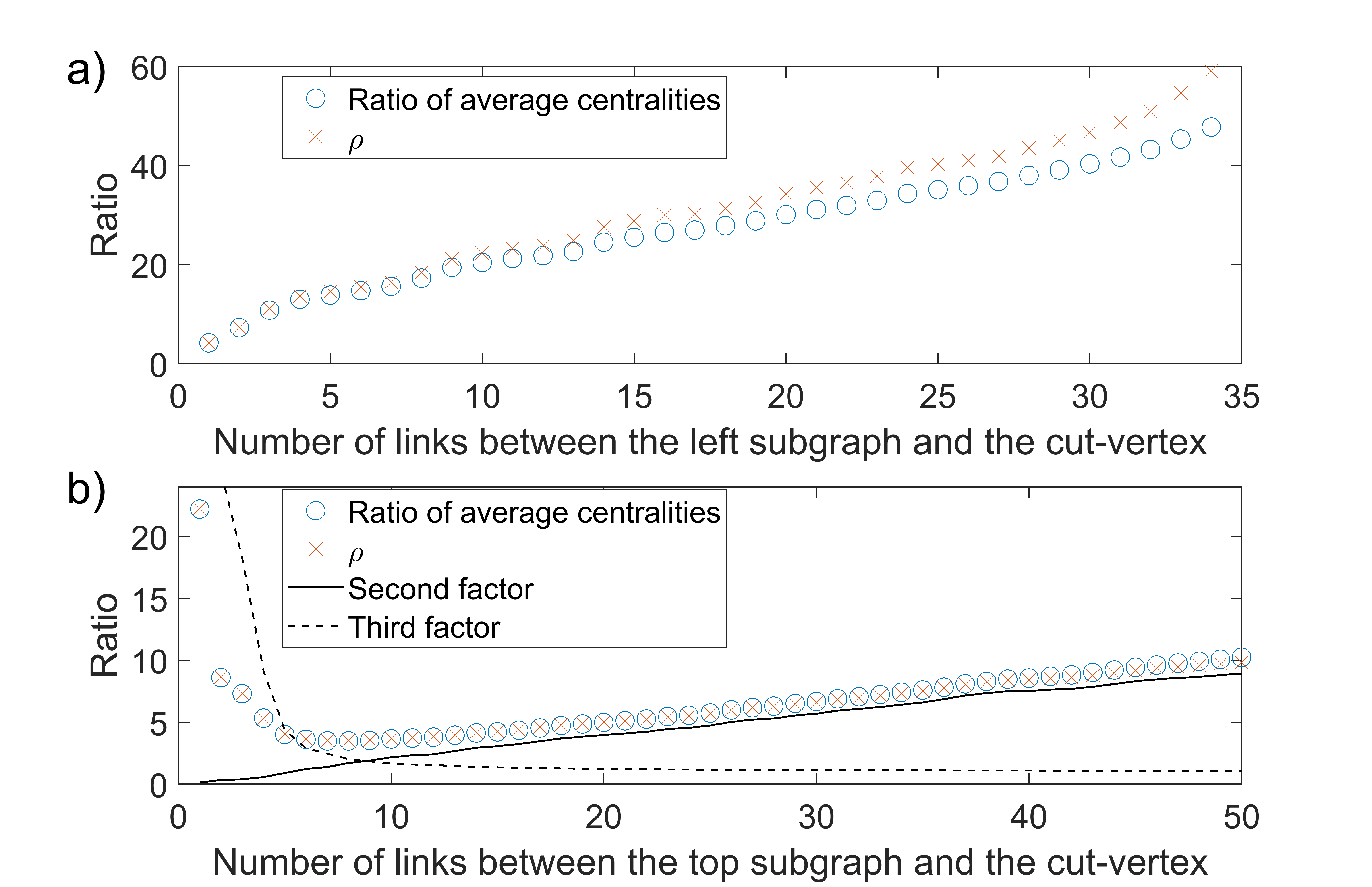}}
\caption{(a) For the network in Fig.~\ref{Karate_twice}, the right subgraph remains connected via node 13 whereas the left subgraph is increasingly connected to the cut vertex, starting with node 1, and then nodes 1 and 2, continuing until all 34 nodes are connected. The impact on the ratio of the average eigenvector centrality (blue circles) as well as on the approximation $\rho$ (red crosses) is shown. (b) For the network in Fig.~\ref{random}, the connectivity of the lower subgraph to the cut vertex remains as before, but the connectivity to the upper subgraph increases by sequentially connecting a new node chosen uniformly at random from the remaining unconnected nodes, except that the first three nodes are chosen to be the same as in Fig.~\ref{random}. The ratio of the average eigenvector centrality (blue circles) and the approximation $\rho$ (red crosses) is shown as well as the contributions of the second (black solid line) and third (black dashed line) factors in Eq.~(\ref{ratio}).}
\label{increasing_connections}
\end{figure}

Since the form of the approximation Eq.~(\ref{ratio}) reduces to just the second factor for the network in Fig.~\ref{Karate_twice}, we conclude that increasing the number of links between the cut vertex and left subgraph will cause the centrality of nodes in $P_1$ to increases with respect to their counterparts in $P_2$. The effect of doing this is illustrated in Fig.~\ref{increasing_connections}(a) for both the actual ratio and the approximation Eq.~(\ref{ratio}). 

In the general case where $P_1$ and $P_2$ are not isomorphic, increasing the number of links to $P_1$ from the cut vertex will still cause the second factor in Eq.~(\ref{ratio}) to increase. However, the value of $\mu$ will also increase (\cite{Grantmacher}, pp. 69-70) which brings in a changing contribution from the third factor in (\ref{ratio}). As $\mu$ increases, this factor will decrease if $\lambda_1>\lambda_2$ and will approach 1 from above, otherwise, aside from the equality case, it increases and approaches 1 from below. In the former case, it is therefore possible that $\rho$ will initially decrease with increasing links, prior to it increasing again. This is illustrated in Fig.~\ref{increasing_connections}(b) by adding connections to the upper subgraph in Fig.~\ref{random}. 

We discussed at the beginning of this section that as $\mu$ gets large with respect to the leading eigenvalue of any given subgraph $P_i$, the centrality of this subgraph approaches the uninformative distribution $\textbf{b}_i$. However, in the case of Fig.~\ref{increasing_connections}(b), we are far from this limit since the principal eigenvalue only increases to $\mu=11.39$ when all fifty nodes in the upper subgraph are connected, which is less than twice the principal eigenvalue of either subgraph. 

The original (single) karate club network also has a cut vertex at node 1 and so for completeness we can consider this. Removal of node 1 fragments the network into three parts: nodes (5,6,7,11,17), node 12, and the remaining 27 nodes. We can determine eigenvector centrality by Eq.~(\ref{EV_cent}) with $m=3$. Identifying partition $P_3$ with node 12, we get $M_3\textbf{b}_3=1$. If we identify partition $P_1$ with nodes (5,6,7,11,17) and their internal connections and $P_2$ with the larger partition, then the ratio of the average eigenvector centrality of nodes in $P_1$ to the average in $P_2$ is 0.4266. The value from Eq.~(\ref{ratio}) is $\rho=0.4612$. Here the first factor is 2.708, the second is 1.272 and the third is 0.1339. There is nothing obviously problematic with the value of this ratio, but its utility in defining the relative importance of the nodes is questionable given our previous observations. 

\section{Localization in networks with a vertex cut set}
\label{4}
We can view the cut vertex as an extreme example of a partitionable graph. This leads to the question of whether the eigenvector centrality of networks which can be partitioned by the removal of a small number of nodes may also exhibit similar localization problems. This can be addressed by generalizing our previous analysis to a vertex cut set. Proceeding by analogy with the cut vertex analysis, we consider an undirected strongly connected network with adjacency matrix $A$ such that the removal of a set of $q$ vertices results in $m$ partitions with adjacency matrices $P_i$ of order $p_i$ by $p_i$ for $i\in\{1,2,\dots,m\}$. We suppose that the internal connections between the $q$ nodes in the vertex cut set are represented by the adjacency matrix $Q$. The matrix $A$ then takes the form
\beq
A=
\left (
\begin{array}{ccccc}
P_1 & 0_{p_1\times p_2} & \cdots & 0_{p_1\times p_m} & B_1     \\
0_{p_2\times p_1} & P_2  & \cdots & 0_{p_2\times p_m} & B_2     \\
\vdots & \vdots &  \ddots & \vdots & \vdots \\
0_{p_m\times p_1} & 0_{p_m\times p_2}  & \cdots & P_m & B_m \\
B^T_1 & B^T_2  &\cdots & B^T_m & Q
\end{array}
\right ),
\nonumber
\eeq
where $B_i$ is a $p_i$ by $q$ matrix denoting the connections from $Q$ to $P_i$. Using the form
\beq
\textbf{u}=\left (
\begin{array}{c}
\textbf{x}_1 \\
\textbf{x}_2 \\
\vdots \\
\textbf{x}_m \\
\textbf{v}
\end{array}
\right )
\nonumber
\eeq
to denote the eigenvector, where now $\textbf{v}$ is a column vector of length $q$ and where $\textbf{x}_i$ is a column vector of length $p_i$, we obtain a generalization of Eq.~(\ref{EV_cent}):
\beq
\textbf{u}\propto\left (
\begin{array}{c}
M_1 B_1\textbf{v}\\
M_2 B_2\textbf{v}  \\
\vdots \\
M_m B_m\textbf{v}  \\
M_Q\left (\sum_{i=1}^m B_i^T\textbf{x}_i\right )
\end{array}
\right ),
\label{EV_cent2}
\eeq
where the matrices $M_i$ are given by Eq.~(\ref{M_defA}) and where
\beq
M_Q=\left (I-\frac{1}{\mu}Q\right )^{-1}.
\nonumber
\eeq
This generalizes our previous analysis by replacing the vector $\textbf{b}_i$ by the vector $B_i\textbf{v}$, and by replacing the scalar $\mu$ in the last element by a vector describing the eigenvector centrality of the vertex cut set. The analysis leading to Eq.~(\ref{ratio}) can be repeated leading to
\beq
\rho=\frac{p_2(\textbf{1}\cdot\textbf{w}_1)}{p_1(\textbf{1}\cdot\textbf{w}_2)}\frac{(B_1\textbf{v}\cdot\textbf{w}_1)}{(B_2\textbf{v}\cdot \textbf{w}_2)}\frac{(\mu-\lambda_2)}{(\mu-\lambda_1)}
\label{ratio2}
\eeq
describing the ratio of the average eigenvector centralities of the first and second partition. As in Eq.~(\ref{ratio}), $\lambda_1$, $\textbf{w}_1$ and $\lambda_2$, $\textbf{w}_2$ are the principal eigenvalues and eigenvectors of partitions $P_1$ and $P_2$, respectively.

\begin{figure}
\centerline{\includegraphics[width=.5\textwidth]{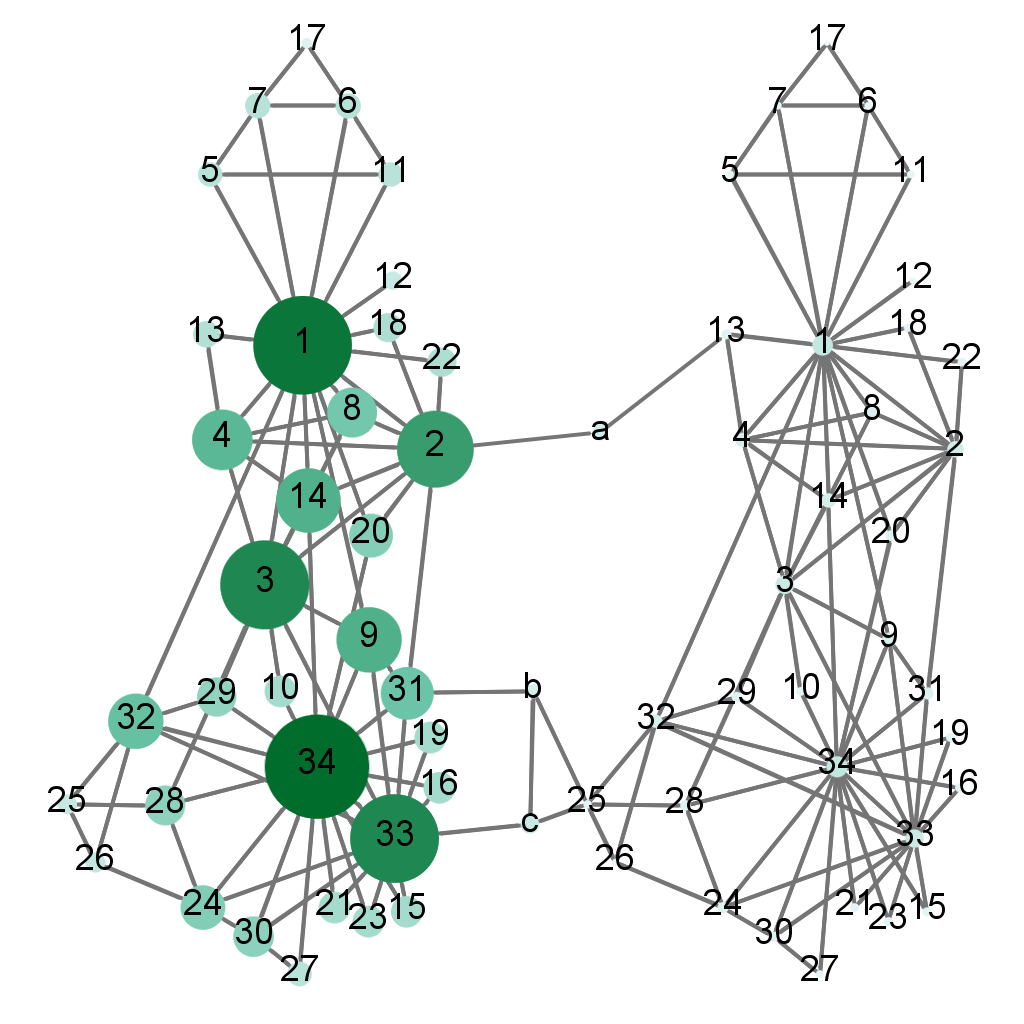}}
\caption{Modification of the double karate network in Fig.~\ref{Karate_twice} to have a vertex cut set of three nodes \{a,b,c\}. The size of the nodes increases with their eigenvector centrality and the color from white through to green is also changing with increasing eigenvector centrality.}
\label{Karate_cut_vertex_set}
\end{figure}

The first factor is the same as for the single cut vertex and the third factor remains unchanged in form and so exhibits the same problems. The second factor is a generalization of the second factor in Eq.~(\ref{ratio}). The main difference is that it depends directly on the eigenvector centrality of the vertex cut set, denoted by $\textbf{v}$, whereas for a single cut vertex we were able to remove this as a common factor. Here, $B_i\textbf{v}$ is a mapping of the centrality $\textbf{v}$ via the linking edges defined by $B_i$ to $P_i$. Only the nodes of $P_i$ that connect directly to the vertex cut set have non-zero entries in $B_i\textbf{v}$ and so this has the same role as $\textbf{b}_i$ in Eq.~(\ref{ratio}). Similarly, the dot product of $B_i\textbf{v}$ with the isolated eigenvector centrality $\textbf{w}_i$ of $P_i$ denotes how much centrality this overlaps with. As in the single cut vertex case, if there are relatively few links between $Q$ and $P_i$, then the centrality of all nodes in $P_i$ can be very dependent on a small change in $B_i$ yielding a potentially large nonlocal impact. This is illustrated in Fig.~\ref{Karate_cut_vertex_set}, 
which represents a modification of the double karate network in Fig.~\ref{Karate_twice} to a slightly larger vertex cut set of three nodes which are chosen to connect to reasonably high degree nodes in the left partition and low degree nodes in the right partition. We see similar behavior to Fig.~\ref{Karate_twice}, where this time the average ratio between corresponding nodes in the left and right graphs (excluding nodes 2,13,25,31,33 which have direct connections to the vertex cut set) is 5.921. This is well-accounted for by $\rho=5.934$.

\section{Types of localization and the nonbacktracking algorithm}
\label{5}
Our analysis has identified three seemingly distinct types of localization which we summarize below.
\begin{itemize}
\item{Type 1:} If a class of network scales with the number of nodes $n$ such that its principal eigenvalue is $\mu(n)$ and if one of the partitions $P_i$ of a vertex cut set has principal eigenvalue $\lambda_i(n)$ and if $\lim_{n\rightarrow\infty}\lambda_i(n)/\mu(n)=0$, then it follows from Eqs.~(\ref{EV_cent2}) and~(\ref{M_defA}) that the proportion of eigenvector centrality allocated to $P_i$ vanishes except for those nodes in the partition connected directly from the vertex cut set. In some circumstances this leads to an unreasonable focusing of centrality. We did not explicitly show this type of localization here, but it has been demonstrated elsewhere (see the discussion at the beginning of Sec.~\ref{3}).
\item{Type 2:} The second factor in Eq.~(\ref{ratio2}) describes a nonlocal impact across the entire of a partition $P_i$ of the choice of nodes connecting it to the vertex cut set. This is particularly apparent when the number of edges connecting $P_i$ to the vertex cut set is low (e.g., Figs.~\ref{Karate_twice} and~\ref{Karate_cut_vertex_set}).
\item{Type 3:}  If the principal eigenvalue of a partition is close to the principal eigenvalue of the full network, then the third factor of Eq.~(\ref{ratio2}) shows how the centrality of this subgraph can become unreasonably high [e.g., Figs.~\ref{random} and~\ref{increasing_connections}(b)].
\end{itemize}

Type 1 localization that is caused by the presence of high-centrality hub nodes has been qualitatively explained on undirected random graphs with a vanishingly small density of short loops in terms of the eigenvalue equation \cite{Martin}. It emerges from the process where a high-centrality hub node passes centrality to its neighbors, but this is then reflected back to the hub via the bidirectional links. This type of backtracking can be avoided by using a modified `nonbacktracking' version of eigenvector centrality \cite{Martin} based on the Hashimoto or nonbacktracking matrix \cite{Hashimoto}\cite{Sodin}\cite{Krzakala}. It is therefore of interest to determine its efficacy on the other types of localization described here. 

If we apply the nonbacktracking variant of eigenvector centrality to the network in Fig.~\ref{Karate_twice}, then we obtain a ratio of corresponding nodes (excluding those immediately connected to the cut vertex) of 1.0002 with standard deviation 0.0020 and so this problem seems to be resolved. However, if we apply it to the network in Fig~\ref{Karate_cut_vertex_set}, we obtain an average ratio of corresponding nodes (excluding those immediately connected to the vertex cut set) of 1.2507 with standard deviation 0.0090 and so the ratio is reduced but the nonlocal influence remains. So it appears that some but not all Type 2 problems can be resolved by this method. 

For the network in Fig.~\ref{random}, the average nonbacktracking centralities in the upper graph are 5.725 times bigger on average than those in the lower graph. This suggests that the localization problems associated with the third factor in Eq.~(\ref{ratio}) remain. To understand these results in more detail, it would be valuable to determine whether an expression similar to Eq.~(\ref{EV_cent2}) could be derived for the nonbacktracking algorithm.

\section{An interpretation of eigenvector centrality}
\label{6}
We have shown that eigenvector centrality can be unreliable due to three different types of localization. 
However, other eigenvector-related centrality measures such as Katz centrality \cite{Katz} and PageRank \cite{BrinPage} are much more robust and do not exhibit the same problems. 

We conclude by arguing that eigenvector centrality can be viewed as an approximation to more robust centralities in the full span of the eigenvectors of the adjacency matrix. In particular, we shall argue this by considering the relationship between Katz centrality and eigenvector centrality.

Katz centrality is defined for a general adjacency matrix $A$ by
\beq
\textbf{x}=M\textbf{1},
\label{Katz}
\eeq
where
\beq
M=\left (I-aA\right )^{-1}
\nonumber
\eeq
and where $a$ is a parameter that we are free to choose within the range $0<a<1/\mu$ \cite{Katz}\cite{Sharkey}. 

The matrix $M$ can be written as a power series in $aA$:
\beq
M=I+aA+a^2A^2+\dots \;.
\label{M_sum}
\eeq
The element $s_{ij}$ of matrix $A^r$ for $r\in \{1,2,\dots \}$ is the number of paths of length $r$ between node $j$ and node $i$. In the original interpretation of this series by Katz, it is supposed that the influence of node $j$ on node $i$ via a path between them reduces by the length of this path according to $a^r$ where $a$ is the `attenuation' on each link, so that shorter paths contribute more. According to Katz, the interpretation of element $M_{ij}$ of matrix $M$ is the influence that node $j$ has on node $i$ due to all possible paths between $j$ and $i$. By performing the sum over $j$ in Eq.~(\ref{Katz}), we determine the total influence of all nodes on node $i$.

Following similar arguments to the derivation of Eq.~(\ref{M_expansion}), Katz centrality can be written in terms of an eigenbasis $\{\textbf{u}_1,\textbf{u}_2,\dots,\textbf{u}_n\}$ (with corresponding eigenvalues $\mu_1,\mu_2,\dots,\mu_n$) of $A$:
\beq
\textbf{x}=\frac{f_1\textbf{u}_1}{1-a\mu_1}+\frac{f_2\textbf{u}_2}{1-a\mu_2}+ \dots + \frac{f_n\textbf{u}_n}{1-a\mu_n},
\nonumber
\eeq
where $f_i$ are the coordinates of vector $\textbf{1}$ in this basis. We can assume that $\mu_1$ is the principal eigenvalue of $A$.

Katz centrality is therefore a vector in the full span of an eigenbasis of $A$. The level of contribution of each eigenvector depends on the parameter $a$. As the parameter gets close to $1/\mu_1$ from below, the first term dominates and, after appropriate normalization, we obtain the convergence of eigenvector and Katz centrality \cite{Bonacich01}\cite{Benzi}. As a result, the same localization problems emerge. However, when we approach the eigenvector centrality limit, each term in the sum Eq.~(\ref{M_sum}) makes a similar contribution to the centrality and the contributions of the later terms converge to each other in size. With endlessly repeated cycles and infinite path lengths, the original process that Katz envisaged \cite{Katz} loses its meaning. 

For lower values of $a$, the attenuation automatically reduces the impact of large paths. At the same time, we gain contributions from the other eigenvectors and so in this sense, Katz centrality can be viewed as a mechanism for assimilating information from all of the eigenvectors of $A$ where $a$ is a tuning parameter to determine the relative magnitude of those contributions. Motivated by this, one useful way of defining the attenuation is $a=1/(\mu_1+\mu_2)$ where $\mu_2$ is the second-largest positive eigenvalue, if it exists. This is bounded by $0.5/\mu_1<a<1/\mu_1$ which is consistent with the value $a=0.85/\mu_1$ used in the related PageRank algorithm \cite{BrinPage}. 

In conclusion, eigenvector centrality can be viewed as the leading contribution to more robust measures in the span of the eigenvectors such as Katz centrality, based on underpinning systems with a clear centrality interpretation \cite{Katz}\cite{Sharkey}. Indeed, we have already argued that there is no sufficient reason why the principal eigenvector of the adjacency matrix should be a reliable centrality measure.

\section{Acknowledgments} The author acknowledges support from the EPSRC, grant No EP/N014499/1.

\end{document}